\documentclass[11pt]{article}

\usepackage[T1]{fontenc}
\usepackage[utf8]{inputenc}
\usepackage{lmodern}
\usepackage[a4paper,margin=1in]{geometry}
\usepackage{amsmath,amssymb,amsthm}
\usepackage{microtype}
\usepackage{needspace}
\usepackage{xcolor}
\usepackage{enumitem}
\usepackage[round,authoryear]{natbib}
\usepackage[colorlinks=true,allcolors=blue!55!black]{hyperref}

\newtheorem{theorem}{Theorem}
\newtheorem{lemma}{Lemma}
\newtheorem{corollary}{Corollary}
\theoremstyle{definition}
\newtheorem{definition}{Definition}
\theoremstyle{plain}
\newcommand{\CEA}{\operatorname{CEA}}

\title{An Enforcement Foundation for Constrained Equal Awards}
\author{Sinan Ertemel\thanks{Department of Economics, Istanbul Technical University, Ma\c{c}ka 34367 Istanbul, T\"urkiye.}}
\date{}

\begin{document}
\maketitle

\begin{abstract}
Standard claims problems take claims as given. We ask how rule selection changes when a certified claim can be augmented by a privately known component that is costly to verify. In a binary environment, equal treatment of equals, claim boundedness, and efficiency impose a sharp lower bound on the gain from exaggeration and hence on the expected auditing needed for truthful reporting to be weakly dominant. The constrained equal awards (CEA) rule attains this bound at every report profile through a correlated audit policy. Conversely, every profilewise audit-minimal rule is CEA. The pointwise result also makes CEA the unique admissible rule that minimizes expected verification cost under every prior over true claims.
\end{abstract}


\section{Introduction}

Claims problems divide an insufficient estate among competing claims. The classical model treats claims as established before allocation \citep{ONeill1982,Thomson2019}. When a documented claim can be augmented by an increment requiring further evidence, the allocation rule also affects verification costs.\footnote{Asbestos trusts use expedited review to assign scheduled claim values from standard medical and exposure criteria; individual review assesses additional claim-specific factors \citep[pp.~15--16, 20--21]{DixonMcGovernCoombe2010}. Individual review can yield higher or lower values, and a payment percentage may scale the scheduled values. This evidentiary distinction motivates the model; the common baseline and single increment are stylizations.}

The canonical claims problem is a triple $(N,E,c)$, where $c_i\ge0$ and $0<E\le\sum_{j\in N}c_j$. A rule chooses awards $x$ satisfying $0\le x_i\le c_i$ and $\sum_{i\in N}x_i=E$. The constrained equal awards rule (CEA) assigns
\[
 \CEA_i(N,E,c)=\min\{c_i,\lambda\},
 \qquad \text{where }\lambda\ge0\text{ solves }\sum_{j\in N}\min\{c_j,\lambda\}=E.
\]
Thus all unsatisfied claimants receive the same award. \citet{Dagan1996} characterizes CEA by \emph{equal treatment of equals} (equal claims receive equal awards), \emph{composition up} (for $E<E'$, allocating $E$ and then $E'-E$ to the residual claims using the same rule yields the allocation for the larger estate $E'$), and \emph{claims-truncation invariance} (the part of a claim that exceeds the estate does not matter). By contrast, the proportional rule assigns $P_i(N,E,c)=Ec_i/\sum_{j\in N}c_j$: it preserves claim ratios and equalizes satisfaction ratios $x_i/c_i$ for positive claims, rather than award levels.

This raises a rule-selection question: among claim-bounded and efficient rules that treat equals equally (henceforth \emph{admissible}), which is cheapest to enforce when eligibility for the increment is privately known? A larger payoff gain from a high report strengthens the incentive to exaggerate and therefore requires more auditing. Verification burden is thus an endogenous property of the rule.

We study a binary environment. Every claimant has the same certified component $L$ and a total claim of either $L$ or $H>L$; eligibility for the increment $H-L$ is privately known, and an audit reveals perfectly whether an asserted increment is genuine. The estate covers all certified components but not all high claims. ``Certified'' describes evidence, not a guaranteed award.

Our main result characterizes CEA. An admissible rule is \emph{profilewise audit-minimal} if no admissible rule can make truthful reporting weakly dominant with fewer expected audits at any report profile. Admissibility imposes a lower bound on the payoff premium from a high report. CEA reaches the resulting audit bound at every profile and is the only rule that does so. The sanction changes the amount of auditing, but not the selected rule.

Recent work derives CEA from restrictions on maximal reasonable claims \citep{FloresSzwagrzakTreibich2025} or immunity to merging manipulations \citep{YinCallejaIzquierdo2026Manipulations}. Strategic approaches instead endogenize claims or bargaining \citep{AshlagiKaragozogluKlaus2012,LiJu2023}. For rationing with single-peaked preferences, \citet{Sprumont1991} characterizes the uniform rule by strategy-proofness, efficiency, and anonymity. Under excess demand it gives the CEA allocation when peaks are interpreted as claims. Under this rule, overstating a peak cannot increase the award of an agent rationed below it; an agent already at her peak does not prefer more. Here claims are entitlements, while preferences increase with the monetary award, including amounts above the true claim. Verification is therefore needed to induce truthful CEA.\footnote{Under excess supply the uniform rule allocates the entire endowment, whereas claim boundedness here leaves any surplus undistributed.} Whereas the costly-verification literature designs allocation mechanisms subject to verification constraints \citep{Townsend1979,BorderSobel1987,BenPorathDekelLipman2014,MylovanovZapechelnyuk2017}, we compare admissible claims rules by the minimum auditing needed to implement each one.

For intuition, let $R$ be the estate remaining after reserving $L$ for every claimant. At a scarce profile with $m$ high reports, claim boundedness caps each low reporter at $L$, so efficiency and equal treatment of equals force each high reporter to receive at least $L+R/m$. A low claimant facing the same $m-1$ high-reporting opponents can therefore gain at least $R/m$ by reporting high. CEA makes both bounds tight.

\section{Model}

Let $N=\{1,\ldots,n\}$, $n\ge2$, be the set of claimants. Fix claim levels $0<L<H$, an estate $E$ with $nL<E<nH$, a per-audit cost $\kappa>0$, and a sanction $p>0$ for a detected false high report. Each claimant $i$ has a true claim $c_i\in\{L,H\}$ and submits a report $r_i\in\{L,H\}$; $c$ and $r$ denote the true and report profiles, both in $\{L,H\}^n$. Let $D\equiv H-L$ be the possible increment and
\[
 R\equiv E-nL\in(0,nD)
\]
the estate remaining after hypothetically covering every certified component. For a profile $r$, let $\mathcal H(r)\equiv\{i\in N:r_i=H\}$ be its set of high reporters, and write $(L,r_{-i})$ for the profile obtained by replacing $i$'s report with $L$.

A \emph{rule} $F$ assigns an award vector $F(r)$ to every report profile $r$ (the estate is fixed and suppressed). It is \emph{admissible} if it is claim bounded ($0\le F_i(r)\le r_i$ for every $i$), efficient ($\sum_{i\in N}F_i(r)=\min\{E,\sum_{i\in N}r_i\}$), and satisfies \emph{equal treatment of equals} (ETE): $r_i=r_j$ implies $F_i(r)=F_j(r)$.\footnote{Anonymity requires $F_{\pi(i)}(\pi r)=F_i(r)$ for every permutation $\pi$, where $(\pi r)_{\pi(i)}=r_i$. Swapping equal reporters leaves $r$ unchanged and therefore implies ETE. CEA satisfies both properties.} Efficiency exhausts the compensation budget whenever claim bounds permit. If $\sum_i r_i<E$, claims are paid in full and the authority retains the surplus. This binary rationing problem restricts to the usual claims domain when total claims are at least $E$.

At a profile $r$ with $m$ high reports, ETE gives high reporters a common award $b^F(r)$ ($m\ge1$) and low reporters a common award $a^F(r)$ ($m<n$). The profile is \emph{nonscarce} if $mD\le R$ and \emph{scarce} otherwise. Claim boundedness and efficiency fully compensate every reporter at nonscarce profiles. At a scarce profile,
\begin{equation}\label{eq:balance}
 (n-m)a^F(r)+mb^F(r)=E,
\end{equation}
where the first term is absent when $m=n$. On this domain CEA depends only on $m$:
\begin{equation}\label{eq:cea}
 a^{\CEA}(r)=L\quad(m<n),
 \qquad
 b^{\CEA}(r)=L+\delta_m^*\quad(m\ge1),
 \qquad\text{where }\ \delta_m^*\equiv\min\left\{D,\frac{R}{m}\right\}.
\end{equation}
Note that $\delta_m^*$ is nonincreasing in $m$, so $\delta_h^*\ge\delta_m^*$ whenever $1\le h\le m$.

We now turn to verification. An \emph{audit policy} $\alpha$ assigns to every report profile $r$ a probability distribution $\alpha(\cdot\mid r)$ over subsets $A\subseteq\mathcal H(r)$. Only high reports are audited, because only the asserted increment requires verification. After observing $r$, the authority draws $A$ and audits its members simultaneously, so $\alpha$ cannot depend on audit outcomes or on $c$. Each audit costs $\kappa$ and reveals the audited claimant's true claim. Audited false high reports are corrected to $L$ before $F$ is applied, and each detected liar pays $p$; the sanction does not augment $E$, and truthful audited claimants are not sanctioned. Claimants are risk neutral and value their award net of any sanction.

Write $q_i^\alpha(r)\equiv\sum_{A\ni i}\alpha(A\mid r)$ for the probability that $i$ is audited. The number of audits at $r$ is the random variable $|A|$, where $A\sim\alpha(\cdot\mid r)$, and its expected value is
\[
 N_\alpha(r)\equiv\sum_{A\subseteq\mathcal H(r)}|A|\,\alpha(A\mid r)=\sum_{i\in\mathcal H(r)}q_i^\alpha(r).
\]
Expected audits depend only on marginals; incentives depend on the joint corrections, which determine the profile to which $F$ is applied.

An audit policy \emph{makes $F$ dominant-strategy incentive compatible (DSIC)} if, for every $i$, true profile $c$, and reports $r_{-i}$ of the others, reporting $c_i$ weakly maximizes $i$'s expected payoff. Let $\mathcal A_p(F)$ denote these policies. For every admissible $F$, auditing every high report is DSIC: exaggerations are corrected and penalized, while truthful high types receive more than $L$ (see the proof of Lemma~\ref{lem:floor}) and low reports yield at most $L$. Hence $\mathcal A_p(F)$ is nonempty. It is closed in a finite product of simplices, so the minimum
\[
 A_F(r)\equiv\min_{\alpha\in\mathcal A_p(F)}N_\alpha(r)
\]
exists. The fixed sanction is suppressed in $A_F$ and subsequent cost notation.

\begin{definition}[Profilewise audit-minimality]\label{def:auditminimal}
An admissible rule $F$ is \emph{profilewise audit-minimal} if $A_F(r)\le A_G(r)$ for every admissible rule $G$ and every $r\in\{L,H\}^n$.
\end{definition}

Fix $r$ with $m\ge1$ high reports and $i\in\mathcal H(r)$. Define
\[
 \Delta_i^F(r)\equiv b^F(r)-a^F(L,r_{-i})
\]
as the \emph{reporting premium}, the gain from reporting high without audits. Suppose $c_i=L$ and $c_{-i}=r_{-i}$, so others' reports are truthful and never corrected. Truth pays $a^F(L,r_{-i})$; exaggeration pays $b^F(r)$ without an audit and $a^F(L,r_{-i})-p$ otherwise. DSIC requires
\[
 (1-q_i^\alpha(r))\,b^F(r)+q_i^\alpha(r)\bigl(a^F(L,r_{-i})-p\bigr)\le a^F(L,r_{-i}).
\]
Equivalently, $q_i^\alpha(r)(\Delta_i^F(r)+p)\ge\Delta_i^F(r)$. Solving at equality for a positive premium gives the threshold
\[
 \widehat q_i^F(r)\equiv\frac{\max\{\Delta_i^F(r),0\}}{\max\{\Delta_i^F(r),0\}+p},
\]
which is zero for a nonpositive premium, since no audit is needed to deter this deviation. Every $\alpha\in\mathcal A_p(F)$ thus satisfies $q_i^\alpha(r)\ge\widehat q_i^F(r)$. This condition is necessary for DSIC but not sufficient, since it considers only truthful opponents. Lemma~\ref{lem:floor} establishes strictly positive premiums. CEA's premium with $m$ high reports is $\delta_m^*$ from \eqref{eq:cea}, giving the marginal audit threshold
\[
 q_m^*\equiv\frac{\delta_m^*}{\delta_m^*+p}.
\]

\section{Results}

\begin{lemma}[Premium floor]\label{lem:floor}
For every admissible $F$, every $r$ with $m\ge1$ high reports, and every $i\in\mathcal H(r)$, $\Delta_i^F(r)\ge\delta_m^*$, with equality for $F=\CEA$.
\end{lemma}

\begin{proof}
If $mD\le R$, both $r$ and $(L,r_{-i})$ are nonscarce, so $b^F(r)=H$, $a^F(L,r_{-i})=L$, and $\Delta_i^F(r)=D=\delta_m^*$. If $mD>R$, claim boundedness gives $a^F(r)\le L$ when $m<n$, and \eqref{eq:balance} yields $mb^F(r)\ge E-(n-m)L=mL+R$; when $m=n$, \eqref{eq:balance} gives $b^F(r)=L+R/n$ directly. Since also $a^F(L,r_{-i})\le L$, we obtain $\Delta_i^F(r)\ge R/m=\delta_m^*$. Equality for CEA follows from \eqref{eq:cea}.
\end{proof}

\begin{lemma}[Audit floor]\label{lem:auditfloor}
For every admissible $F$ and every $r$ with $m\ge1$ high reports,
\[
 A_F(r)\ \ge\ \sum_{i\in\mathcal H(r)}\widehat q_i^F(r)\ \ge\ m\,q_m^*.
\]
\end{lemma}

\begin{proof}
Every $\alpha\in\mathcal A_p(F)$ satisfies $N_\alpha(r)=\sum_{i\in\mathcal H(r)}q_i^\alpha(r)\ge\sum_{i\in\mathcal H(r)}\widehat q_i^F(r)$; minimizing over $\alpha$ gives the first inequality. The second follows from Lemma~\ref{lem:floor}, since $z\mapsto z/(z+p)$ is strictly increasing on $z\ge0$.
\end{proof}

\begin{theorem}[Audit-minimality characterization of CEA]\label{thm:main}
The audit policy $\alpha^*$ defined by $\alpha^*(\emptyset\mid r)=1$ if $\mathcal H(r)=\emptyset$ and, if $r$ has $m\ge1$ high reports,
\[
 \alpha^*(\mathcal H(r)\mid r)=q_m^*,\qquad \alpha^*(\emptyset\mid r)=1-q_m^*,
\]
makes CEA DSIC and attains $A_{\CEA}(r)=m\,q_m^*$ at every profile $r$ with $m\ge1$ high reports, with $A_{\CEA}(r)=0$ at the all-low profile. Moreover, for an admissible rule $F$, the following are equivalent:
\begin{enumerate}[label=(\alph*)]
 \item\label{it:cea} $F=\CEA$;
 \item\label{it:delta} $\Delta_i^F(r)=\delta_m^*$ for every $r$ with $m\ge1$ high reports and every $i\in\mathcal H(r)$;
 \item\label{it:min} $F$ is profilewise audit-minimal.
\end{enumerate}
\end{theorem}

\begin{proof}
Fix $i$ and the others' true claims and reports, and let $m$ count high reports when $i$ reports $H$. Under $\alpha^*$, each high reporter is audited with probability $q_m^*$, so $N_{\alpha^*}(r)=m\,q_m^*$. Reporting $L$ always yields $L$ under CEA.

Let $c_i=L$. Reporting $H$ yields $L+\delta_m^*$ if nobody is audited. Otherwise $i$ is detected, corrected, and receives $L-p$. The expected payoff is $(1-q_m^*)(L+\delta_m^*)+q_m^*(L-p)=L$, so truthful reporting is weakly optimal.

Let $c_i=H$. Reporting $H$ yields $L+\delta_m^*>L$ if nobody is audited. Otherwise $i$ is confirmed while false high reports of others are corrected, leaving $h$ high reports with $1\le h\le m$; $i$ receives $L+\delta_h^*\ge L+\delta_m^*>L$ because $\delta^*$ is nonincreasing. Reporting $H$ is strictly optimal. Hence $\alpha^*\in\mathcal A_p(\CEA)$, and with Lemma~\ref{lem:auditfloor}, $A_{\CEA}(r)=m\,q_m^*$; at the all-low profile, $A_{\CEA}(r)=0$.

\emph{\ref{it:cea}$\Rightarrow$\ref{it:delta}} is the equality case of Lemma~\ref{lem:floor}.

\emph{\ref{it:delta}$\Rightarrow$\ref{it:cea}.} Nonscarce profiles are fixed by admissibility. Let $r$ be scarce with $m$ high reports and $i\in\mathcal H(r)$. The proof of Lemma~\ref{lem:floor} gives
\[
 \Delta_i^F(r)=b^F(r)-a^F(L,r_{-i})\ \ge\ L+\frac{R}{m}-a^F(L,r_{-i})\ \ge\ \frac{R}{m}.
\]
By \ref{it:delta}, $\Delta_i^F(r)=\delta_m^*=R/m$, so both inequalities bind and $b^F(r)=L+R/m$. If $m<n$, \eqref{eq:balance} then gives $a^F(r)=L$. Every scarce profile's awards thus coincide with \eqref{eq:cea}, so $F=\CEA$.

\emph{\ref{it:cea}$\Rightarrow$\ref{it:min}} follows from $A_{\CEA}(r)=m\,q_m^*$ and Lemma~\ref{lem:auditfloor}.

\emph{\ref{it:min}$\Rightarrow$\ref{it:delta}.} If $F$ is profilewise audit-minimal, then for every $r$ with $m\ge1$ high reports,
\[
 m\,q_m^*\ \le\ \sum_{i\in\mathcal H(r)}\widehat q_i^F(r)\ \le\ A_F(r)\ \le\ A_{\CEA}(r)=m\,q_m^*.
\]
Since each $\widehat q_i^F(r)\ge q_m^*$ by Lemma~\ref{lem:floor} and the monotonicity of $z/(z+p)$, each equals $q_m^*>0$. Strict monotonicity of $z/(z+p)$ then gives $\Delta_i^F(r)=\delta_m^*$.
\end{proof}

The joint audit distribution matters for attaining the bound. Let $n=2$, $L=1$, $H=3$, $E=4$, and $p=1$, so $q_2^*=1/2$. Fix a claimant, suppose both true claims are low, and let the opponent report $H$. Reporting $L$ pays $1$. If both report $H$ and each is audited independently with probability $1/2$, the four equally likely outcomes (neither audited, only the opponent, only the claimant, both) give a claimant net payoffs $2,3,0,0$: catching only the opponent raises the unaudited liar's award to $3$. Exaggeration therefore pays $5/4>1$. Under $\alpha^*$, both are audited or neither is, and exaggeration pays $(0+2)/2=1$. Correlation thus allows the minimum marginal audit probabilities to deter exaggeration even when others lie. Independent audits at higher rates, including full auditing, can still provide DSIC.

The proportional benchmark also illustrates the strictness of the bound. At a scarce mixed profile $r$ (one with $1\le m<n$ high reports), let $\rho_m\equiv E/(nL+mD)<1$. Proportional awards are $a^P(r)=\rho_mL<L$ and $b^P(r)=\rho_mH$. By efficiency, $b^P(r)>L+R/m$; since $a^P(L,r_{-i})\le L$, it follows that $\Delta_i^P(r)>R/m$ for every $i\in\mathcal H(r)$. Proportional therefore requires strictly more auditing at that profile. If no mixed profile is scarce, its full-compensation extension and CEA coincide on the binary domain.

The sanction changes required auditing, while CEA remains uniquely optimal for every finite $p>0$, including $p>H$. Such penalties are feasible in the model because net payoffs may be negative. Since $q_m^*\to0$ as $p\to\infty$, institutional restrictions justify taking $p$ as fixed or bounded \citep{BorderSobel1987,MylovanovZapechelnyuk2017}. If the authority chooses $0<p\le\bar p<\infty$ without a cost of setting the sanction, $\mathcal A_p(F)$ expands with $p$, so choosing $\bar p$ minimizes audits. A cost incurred only when a sanction is imposed is zero along the truthful equilibrium used to evaluate costs. An ex ante cost of establishing a sanction may favor a smaller $p$. If this cost depends only on $p$, replacing any rule by CEA at the same positive sanction weakly lowers total expected cost. Thus any attained joint optimum can be achieved with CEA.\footnote{If sanctions cannot exceed the corrected award, put $s_i=\min\{p,a^F(L,r_{-i})\}$ and $s^*=\min\{p,L\}>0$. The necessary audit probability is $\Delta_i^F(r)/(\Delta_i^F(r)+s_i)\ge\Delta_i^F(r)/(\Delta_i^F(r)+s^*)\ge\delta_m^*/(\delta_m^*+s^*)$. CEA attains this bound with $\alpha^*$ using $s^*$ in place of $p$. Equality forces the original premium equality, so uniqueness also survives this award-based sanction cap.}

We close with expected verification cost. For a prior $\mu$ over true profiles $\{L,H\}^n$, the minimum expected verification cost of implementing $F$ truthfully is
\[
 C_F(\mu)\equiv\kappa\min_{\alpha\in\mathcal A_p(F)}\mathbb E_{c\sim\mu}\bigl[N_\alpha(c)\bigr].
\]
Costs are evaluated at truthful reports $r=c$. Under $\alpha^*$, low types are indifferent between both reports, so truthfulness need not be the unique equilibrium. The prior may assign all its weight to nonscarce profiles, where every admissible rule pays claims in full.

\begin{corollary}[Prior-free verification cost]\label{cor:cost}
For every prior $\mu$ and every admissible rule $F$, $C_{\CEA}(\mu)\le C_F(\mu)$. Moreover, an admissible rule that minimizes expected verification cost for every prior is CEA.
\end{corollary}

\begin{proof}
For every $\alpha\in\mathcal A_p(F)$ and every $c$, $N_\alpha(c)\ge A_F(c)\ge A_{\CEA}(c)=N_{\alpha^*}(c)$ by the definition of $A_F$ and Theorem~\ref{thm:main}. At the all-low profile nobody reports high, so every policy audits nobody and all these terms are zero. Taking expectations under $\mu$ and minimizing over $\alpha$ gives $C_F(\mu)\ge\kappa\,\mathbb E_{c\sim\mu}[N_{\alpha^*}(c)]= C_{\CEA}(\mu)$. For uniqueness, let $F$ minimize cost for every prior and take $\mu$ degenerate on an arbitrary profile $c$. Then $\kappa A_F(c)=C_F(\mu)\le C_{\CEA}(\mu)=\kappa A_{\CEA}(c)$, and $A_{\CEA}(c)\le A_G(c)$ for every admissible $G$ by Theorem~\ref{thm:main}. As $c$ is arbitrary, $F$ is profilewise audit-minimal, and Theorem~\ref{thm:main} gives $F=\CEA$.
\end{proof}

\section{Conclusion}

Costly verification changes the comparison of claims rules because a rule determines not only how the estate is divided but also how profitable it is to exaggerate. In the binary environment, admissibility imposes a lower bound on this reporting gain. CEA uniquely attains the bound simultaneously at every profile and, together with a correlated audit policy, makes truthful reporting weakly dominant with the fewest expected audits. The contrast with proportional division identifies the source of the result: protecting each certified component leaves the smallest possible premium for a high report. Because the bound is pointwise, the same argument delivers verification-cost optimality under every prior.

The characterization is deliberately tied to perfect, one-sided, simultaneous verification. Richer claim spaces, heterogeneous evidence or sanctions, and adaptive or imperfect audits may select different rules.

\Needspace{7\baselineskip}
\section*{Declaration of generative AI and AI-assisted technologies in manuscript preparation}
ChatGPT (OpenAI) was used to assist with exposition and checks of arguments, notation, and references. The author reviewed and revised the resulting material and takes responsibility for the article.

\bibliographystyle{plainnat}
\bibliography{references}
\end{document}